\documentclass[conference]{IEEEtran}
\IEEEoverridecommandlockouts
\usepackage{multirow}
\usepackage{boldline,multirow}
\usepackage{cite}
\usepackage{textcomp}
\usepackage{xcolor}
\usepackage{booktabs}
\usepackage{tikz}
\usepackage{array} 
\usepackage{subcaption}
\usepackage{mathtools}
\usepackage[normalem]{ulem}
\usepackage[english]{babel}
\usepackage{amsthm}
\usepackage{amsmath,amssymb,amsfonts}
\usepackage{graphicx,color}
\usepackage{textcomp}
\usepackage{algorithm}
\usepackage{algpseudocode}
\usepackage{placeins}
\usepackage{adjustbox}
\usepackage{bm}
\usepackage{multirow}
\usepackage{threeparttable}
\usepackage{makecell}
\usepackage{bbding}

\usepackage{balance}
\usepackage{textcomp}
\usepackage[T1]{fontenc}
\usepackage[left=1.62cm,right=1.62cm,top=1.9cm]{geometry}
\useunder{\uline}{\ul}{}

\newcolumntype{C}[1]{>{\centering\arraybackslash}p{#1}}
\def\BibTeX{{\rm B\kern-.05em{\sc i\kern-.025em b}\kern-.08em
    T\kern-.1667em\lower.7ex\hbox{E}\kern-.125emX}}
\begin{document}

    \title{QEA: An Accelerator for Quantum Circuit Simulation with Resources Efficiency and Flexibility

}

\author{
	\IEEEauthorblockN{
    Van Duy Tran\textsuperscript{1},
    Tuan Hai Vu\textsuperscript{1}, 
    Vu Trung Duong Le\textsuperscript{1}, 
    Hoai Luan Pham\textsuperscript{1}, 
    and Yasuhiko Nakashima\textsuperscript{1}}
	\IEEEauthorblockA{
    \textsuperscript{1} Nara Institute of Science and Technology, 8916–5 Takayama-cho, Ikoma, Nara 630-0192, Japan.\\}
}

\maketitle

\begin{abstract}
    The area of quantum circuit simulation has attracted a lot of attention in recent years. However, due to the exponentially increasing computational costs, assessing and validating these models on large datasets poses significant obstacles. Despite plenty of research in quantum simulation, issues such as memory management, system adaptability, and execution efficiency remain unresolved. In this study, we introduce QEA, a state vector-based hardware accelerator that overcomes these difficulties with four key improvements: optimized memory allocation management, open PE, flexible ALU, and simplified CX swapper. To evaluate QEA's capabilities, we implemented and evaluated it on the AMD Alveo U280 board, which uses only 0.534 W of power. Experimental results show that QEA is extremely flexible, supporting a wide range of quantum circuits, has excellent fidelity, making it appropriate for standard quantum emulators, and outperforms powerful CPUs and related works up to 153.16$\times$ better in terms of normalized gate speed. This study has considerable potential as a useful approach for quantum emulators in future works.
\end{abstract}


\section{Introduction}
    Quantum algorithms have the potential to accelerate optimization, data processing, and feature extraction, which are essential in quantum computing \cite{PRXQuantum.3.030101}. However, development in this field is constrained by the restricted availability and capability of quantum hardware, requiring fast quantum simulation systems. As the number of qubits (\#Qubits) increases, typical state-vector simulations such as naïve matrix multiplication (MM) become inefficient due to exponentially growing the \#Qubits. Various approaches have been explored to address this challenge, including high-performance computing (HPC) systems that offer speed but demand power-intensive hardware such as DGX A100 and H100 \cite{wang2024queenquickscalablecomprehensive}, as well as software optimizations that minimize redundant operations but remain ineffective at large scale \cite{ bergholm2022pennylaneautomaticdifferentiationhybrid}. Alternative simulation techniques, such as tensor networks, the Heisenberg picture, and density matrices, as well as methods other than state-vector simulations, reduce the computational complexity to polynomial or linear time but are limited to specific circuit types. All the above works focus on general-purpose emulators or some fixed quantum algorithms, highlighting the need for more scalable and adaptable quantum simulation approaches.

    Recently, substantial research has concentrated on designing quantum simulation hardware platforms (quantum emulators) employing Field-Programmable Gate Arrays (FPGAs), as a potential solution \cite{silva2017fpga, mahmud2020efficient, hong2022quantum, waidyasooriya2022scalable, hieu2024, liang2024pcq}. This approach provides advantages in terms of programmability and practicality, as well as great performance and efficiency at lower costs. Existing studies, however, focus on specific applications with a restricted number of supported qubits, such as the Quantum Fourier Transform (QFT), Quantum Haar Transform (QHT), and Quantum Support Vector Machine (QSVM). As a result, their implementations lack the versatility required to enable a wider range of applications and more complicated quantum tasks. Furthermore, contemporary FPGA-based quantum emulators frequently encounter issues such as inefficient memory management, inefficient resource use, and slow execution speed, which are mainly due to non-optimized hardware designs.

    To address these problems, we offer a hardware architecture named Quantum Emulation Accelerator (QEA) intended to support a wide range of applications with a high \#Qubits while maintaining both flexibility and high performance at low hardware resources. Our approach combines hardware and software enhancements to efficiently simulate quantum circuits. On the software side, we still follow the state-vector simulator but enhance the performance of the overall process via gate-group-and-fusion and compact tensor product (TP) techniques. On the hardware side, we propose four key approaches, including: 
        \begin{itemize}
            \item {Optimized memory allocation management: A strategy for efficient memory utilization.}
            \item {Open processing element (PE): A processing element designed to allow data sharing among four processing elements, reducing execution time.}
            \item {Flexible arithmetic logic unit (ALU): A computation unit capable of switching between different operational modes to optimize data processing.}
            \item {Simplified CX swapper: A module designed for efficient computation of the CX gate.}
        \end{itemize}

    To verify and evaluate QEA's performance, we implement it on the AMD Alveo U280 board and analyze its operational frequency, flexibility, and execution time.
    

        \begin{figure}[t]
            \centering
            \includegraphics[width=0.4\textwidth]{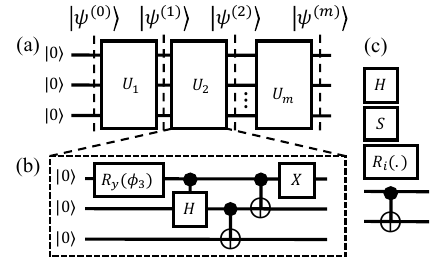}
            \caption{(a) Quantum circuit as unitary operator's chain (b) An example 3-qubit W state circuit \cite{10413647} (c) Universal gate set $\{H, S, R_i(.), CX\}$ with $i\in\{x,y,z\}$ \cite{PhysRevA.70.052328}.}
            \label{fig:overview}  
            \vspace{-5mm}
        \end{figure}
\section{Background knowledge} \label{sec:background}
    \subsection{Quantum simulation algorithm} \label{sec:background:qc_alg}

        A quantum circuit can be simulated by a series of gates that act on a reference state $|\psi^{(0)}\rangle$ as $|\psi^{(m)}\rangle = \mathcal{U}|\psi^{(0)}\rangle$, where $|\psi^{(m)}\rangle$ and $\mathcal{U}(\bm\theta)$ are target state and quantum operator, respectively. This quantum operator can be decomposed into MM between sub-operators but also into $\hat{m}$ gates (considering only single-qubit gates and two-qubit gates) application. These gates $\{g_j\}$ can be applied on $|\psi^{(0)}\rangle$ sequentially to achieve the same final state $|\psi^{(m)}\rangle$:
        
        \begin{align}
            \mathcal{U} \equiv \prod_{j=1}^m U_j =(\prod_{j=1}^{\hat{m}}g_{j}),\;
            |\psi^{(j+1)}\rangle=U_j|\psi^{(j)}\rangle
            \label{eq:basic}
        \end{align}
        
        with the state vector can be viewed as $2^n$-dimensional complex array $\psi=\left[\alpha_0\;\alpha_1\;\ldots\;\alpha_{2^N-1}\right]$ and $\{\alpha_j\}$ is called amplitudes of state vector.

    \subsection{Quantum gates} \label{sec:background:quantum_gate}
        In~\eqref{eq:basic}, quantum gates are fundamental operations applied to qubits, represented as unitary matrices. In this work, we are considering the gate set Clifford + $R_i$ with $i\in\{x,y,z\}$, which is known as a universal gate set followed by Gottesman-Knill theorem \cite{PhysRevA.70.052328}, any $g_j$ will belong to the set $\{H, S, R_x(.), R_y(.), R_z(.), CX\}$ as Fig.~\ref{fig:overview} (c). This set is denoted as fixed gates $\{H,S,CX\}$, parameterized gates $\{R_x(.), R_y(.), R_z(.)\}$, sparse gates $\{S, R_z(.)\}$ and dense gates $\{H, CX, R_x(.), R_y(.)\}$ depend on its matrix representation. While $H, S$ operator on one-qubit, $CX$ applies a NOT operation to a target qubit conditioned on a control qubit.
        
        By using Clifford + $R_i$, we can break down into any low-level or build up any high-level gates, such as control rotation $CR_x(.)$ can be decomposed into $\{S, CX, R_x(.), R_y(.), R_z(.)\}$. This process can be conducted by a transpiler \cite{RAKYTA2024112756}, the transpiled state can recover the original state by $|\psi_{\text{tran}}\rangle=e^{i\phi}|\psi_{\text{origin}}\rangle$ where $\phi$ is the global phase born from the transpilation process.
    
    \subsection{Challenge}
    \label{sec:background:challenge}

    The challenge when emulating~\eqref{eq:basic} for general purpose is the exponential scaling of both computation time and resources based on \#Qubits. In detail, $\{U_j\}$ and $\psi$ can be represented as $N\times N$ complex matrix ($N = 2^n)$ and $N$-dimensional vector, respectively. Take example, storing such one matrix in case $50$ qubits require at least $8\times 10^{31}\;\text{bit} \approx 10^{22}$ (GB) for single precision. $50$ qubits state would take $7.2\times 10^{16} \text{bit} \approx 9 \times 10^{6}$ (GB). Such huge required resources are impossible for more \#Qubits even with the supercomputer.  
    In almost all simulators, only $\psi$ is being saved during the whole process. In QEA, $\{U_j\}$ can be applied directly to $|\psi\rangle$ without construction, the same properties as the wave-function approach \cite{qsun}.

\section{Proposed Hardware Architecture}\label{sec:proposed}
    \subsection{Overview} \label{sec:proposed:overview}
        \begin{figure}[t]
            \centering
            \includegraphics[width=0.49\textwidth]{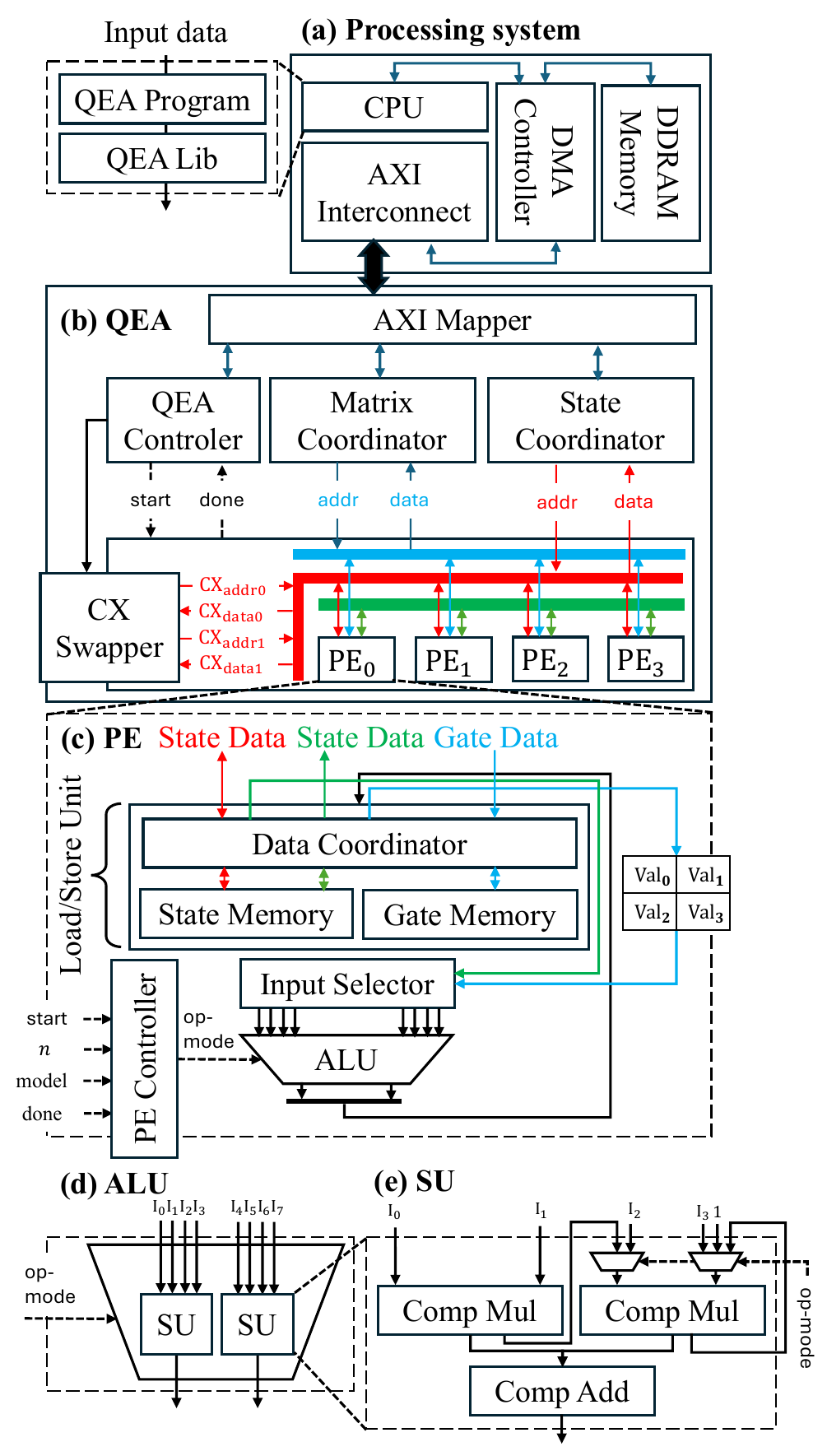}
            \caption{The detail of the QEA: (a) The processing system, (b) QEA core, (c) Processing Element (PE), (d) Arithmetic Logic Unit, (e) Special Unit (SU)}
            \label{fig:hardware_overview}  
            \vspace{-5mm}
        \end{figure}

        Fig.~\ref{fig:hardware_overview} (a-c) shows an overview of QEA, which consists of two primary components: the processing system (PS) and QEA. In the processing system, the QEA library allows users to create quantum programs and generate context data for execution on QEA. Direct Memory Access (DMA) is used to efficiently move large amounts of data between the user space and the PL, with a 256-bit AXI bus optimized for data communication. QEA, the main part of the architecture is composed of six important components:
            \begin{itemize}
                \item {AXI Mapper: Manages the transfer of data between PS and QEA to ensure proper synchronization.}
                \item {QEA Controller: Controls execution management within the accelerator.}
                \item {Matrix Coordinator and State Coordinator: Assist the AXI Mapper by routing data to the state vector and gate context memories in each PE.}
                \item {The CX Swapper: Performs the computation of a CX gate and a state vector (described in Section~\ref{sec:proposed:simp_CX}).}
                \item {The Processing Element Array (PEA): Consists of four PEs that perform computations to update the state vector when sparse and dense gates are applied.}
            \end{itemize}


    \subsection{Optimized Memory Allocation Management} \label{sec:proposed:mem_alloc}
        \begin{algorithm}[t]
            \caption{Matrix multiplication for a gate $g$ at $j$ qubit}
            \begin{algorithmic}[1]
                \Require $n$, $\psi$, $g$, $j$, $u$ (matrix representation of $g$), $N\gets 2^n$
                \State $\overline{g} \gets 1 \ll (n - (j + 1))$ \Comment{Length of sub-generated group}
                \State $f \gets 1$ \Comment{Flag variable}
                \For{$i = 0$ to $N-1$}
                    \If{$f = 1$}
                        \If{$type(g) == 0$} 
                            \State $\psi[i] \gets u[0][0]\times\psi[i]$ \Comment{Sparse gate}
                        \Else $\;\psi[i] \gets u[0][0]\times\psi[i] + u[0][1]\times\psi[i+\overline{g}]$ 
                        \EndIf
                    \Else
                        \If{$type(g) == 1$}
                            \State $\psi[i] \gets u[1][1]\times\psi[i]$  \Comment{Sparse gate}
                        \Else  $\;\psi[i] \gets u[1][0]\times\psi[i-\overline{g}] + u[1][1]\times\psi[i]$ 
                        \EndIf
                    \EndIf

                    \If{$i\mod \overline{g} = \overline{g}-1$} $f \gets \lnot f$
                    \EndIf
                \EndFor
                \State \Return $\psi$
            \end{algorithmic}
            \label{alg:matrix_mul}
            
        \end{algorithm}

        Efficient memory allocation plays an essential role in quantum circuit simulation hardware design. As described in Section~\ref{sec:background:challenge}, the exponential increase in \#Qubits greatly raises memory demands. The accelerator in \cite{hieu2024} uses numerous memory units for gate context, state vector, matrix data, and next state vector storage, resulting in significant memory overhead. Furthermore, the usage of $2\times2$ gate matrices to generate a $2^n \times 2^n$ matrix offers no computational speed-up, leading to inefficient memory utilization. To overcome these limitations, QEA uses efficient memory allocation management, as shown in Fig.~\ref{fig:hardware_overview} (b-d). QEA uses a single global memory for gate context storage and two local memories for the state vector (State Memory) and gate data (Gate Memory) in each PE. The Matrix Coordinator delivers 2$\times$2 gate matrix data directly to each Processing Element (PE) during quantum gate operations (except the CX gate), reducing the amount of gate data stored by PEs. Furthermore, the current state vector is distributed equally among all PEs, while the next state vector is directly updated to the current state vector, maximizing memory efficiency while maintaining computational speed.
        
        \begin{algorithm}[t]
            \caption{CX gate application on state vector}
            \begin{algorithmic}[1]
                \Require $n, \psi, j, k, N \gets 2^n$
                \For{$i = 0$ to $N-1$}
                    \State $\overline{i}_2 \gets \text{bin}(i,n)$ \Comment{Binary representation of $i$ in $n$ bits}
                    \If{$\overline{i}_2[j] = 1$}
                        \State $j \gets i \oplus (1 \ll k)$
                        \State $\text{swap}(\psi[i], \psi[j])$
                    \EndIf
                \EndFor
                \State \Return $\psi$
            \end{algorithmic}
            \label{alg:cx}
        \end{algorithm}

        Besides memory allocation, a fundamental problem is guaranteeing efficient data storage and computation when applying a $2\times2$ gate matrix to a $2^n$-sized state vector. \cite{hieu2024} suggests a Coordinate (COO) format for storing row/column indices and real and imaginary values, but it increases memory requirements. Because indices may be dynamically generated during execution (Algorithm~\ref{alg:matrix_mul}), QEA uses a simpler format that stores only real and imaginary components, avoiding redundant data while retaining correctness. Furthermore, QEA employs a fixed-point 32-bit representation (2-bit integer, 30-bit fractional portion), resulting in faster computations than floating-point arithmetic while maintaining precision in quantum circuit simulations.
        
    \subsection{Open Processing Element} \label{sec:proposed:qpe}
        In a quantum circuit, data flow management is also essential in data computation and memory allocation, especially when multiple PEs are used to parallelize computing. As \#Qubits grows, data access hazards occur, possibly affecting global execution time due to the distribution of state data among PEs. To reduce data hazards, each PE in the PEA allows for state data transfer across PEs. This approach enables state data to be shared between PEs, ensuring concurrent access for both the present PE and other PEs that require the same data. As a result, data access management is improved, and data access time decreases while overcoming the throughput constraints of State Memory within each PE. Fig.~\ref{fig:hardware_overview} (c) and (d) depict the mechanism in detail. Furthermore, because only one quantum gate is used in each phase, gate data can be previously loaded from Gate Memory, minimizing access latency and increasing overall quantum circuit simulation speed.

    \subsection{Flexible Arithmetic Logic Unit} \label{sec:proposed:flex_ALU}
        Besides the CX gate, there are two types of gates: sparse and dense gates. Sparse gates, which have fewer elements than dense gates, can be executed in around half the time, making them faster. Thus, creating an ALU that can switch between sparse and dense gate operations is critical for dramatically reducing execution time. To solve this, QEA integrates an ALU, as shown in Fig. \ref{fig:hardware_overview} (e) and (f), which allows switching between sparse and dense gate computations. This ALU utilizes two complex multiplication operations and one complex addition operation within each Special Unit (SU). Furthermore, to optimize data access performance, the ALU has two SUs, ensuring efficient utilization of the State Memory.

    \subsection{Simplified CX Swapper} \label{sec:proposed:simp_CX}
        The two-qubit structure of the CX gate makes it a relatively difficult operation in a quantum circuit. When applying a CX gate to a state vector, the conventional method usually involves creating a full $n \times n$ CX matrix based on the state vector of $n$ quantum bits and then performing matrix multiplication with the state vector. However, this approach is time-consuming because it requires generating the complete CX matrix and carrying out the multiplication. Fortunately, the main purpose of the CX gate is to switch values inside the state vector. Therefore, as shown in Algorithm.~\ref{alg:cx}, a swap technique can handle the CX gate efficiently \cite{horii2021optimization}. 

        \begin{figure}[t]
            \centering
            \includegraphics[width=0.47\textwidth]{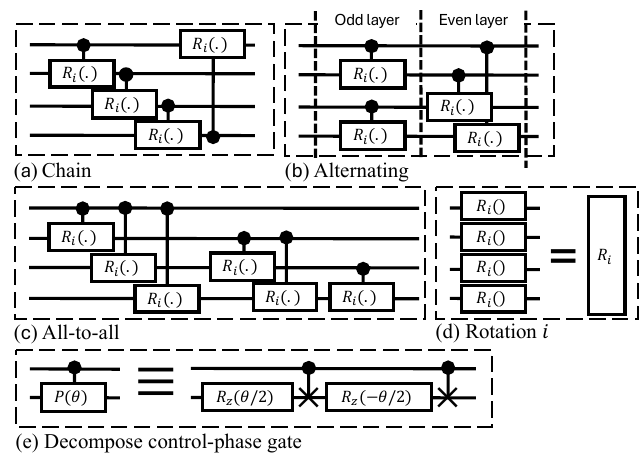}
            \caption{Example of 4-qubit topologies (a-d) and (e) Decomposing of $CP(\theta)$ as $(I\otimes R_z(\theta/2))CX((I\otimes R_z(-\theta/2)))CX$.}
            \label{fig:topology}  
        \end{figure}
\section{Verification and Evaluation} \label{sec:ver_and_eval}
    \subsection{Implementation and Verification} \label{sec:ver_and_eval:impl_and_ver}
        The Quantum Execution Architecture (QEA) was implemented and verified on the 16nm Alveo U280 FPGA board (Fig.~\ref{fig:hardware_overview}) using the Vivado 2020.2 tool. The QEA requires 14,773 lookup tables (LUTs), 7,894 flip-flops (FFs), 578 Block RAMs (BRAMs), and 256 digital signal processors (DSPs) while consuming only 0.543 W of power. With these resources, the QEA can best support quantum circuits of 3 to 17 (due to the limitation of BRAM on Alveo U280) and can be extended for higher \#Qubits.


    \subsection{Benchmarking setting} \label{sec:ver_and_eval:benchmark_cpu}
         
        To prove the generality of QEA, we evaluated 19 circuit templates \cite{expressibility}, each labeled with a circuit ID from 1 to 19. The full circuit representation is presented in Fig. ~\ref{fig:topology}. All circuits are constructed from a combination of topologies, including chain, alternating, all-to-all \cite{PRXQuantum.2.040309}, and rotation; as shown in Fig.~\ref{fig:topology}. In circuit design, the chain topology arranges qubits in a linear sequence, where each qubit is coupled only to its immediate neighbors. The alternating topology extends this by introducing a bipartite structure, connecting qubits in two subsets with edges between them. In contrast, the all-to-all topology allows every qubit to interact directly with every other, forming a complete graph. The final, rotation $i$ takes a role as parameter holder and provides no entanglement.

        For comparison with other works, Quantum Fourier Transform (QFT) \cite{coppersmith2002approximatefouriertransformuseful} is conducted. It is an important component in many quantum algorithms, such as phase estimation, order-finding, and factoring. QFT circuit is utilized from Hadamard ($H$), controlled phase ($CP(\theta)$) is decomposed as Fig.~\ref{fig:topology} (e), and SWAP gates equivalent to three sequential $CX$ gates. These gates transform on an orthonormal basis $\{|0\rangle,\ldots,|N-1\rangle\}$ with the below action on the basis state $|j\rangle\rightarrow\frac{1}{\sqrt{N}}\sum_{k=0}^{N-1}e^{2\pi ijk/N}$. We apply the QFT on the zero state $|\bm{0}\rangle$, which should return the exact equal superposition of all possible $n$-qubit states, means $\alpha_j=\alpha_k \;\forall j,k\in[0,N)$.
        
        \begin{figure}[t]
            \centering
            \includegraphics[width=0.47\textwidth]{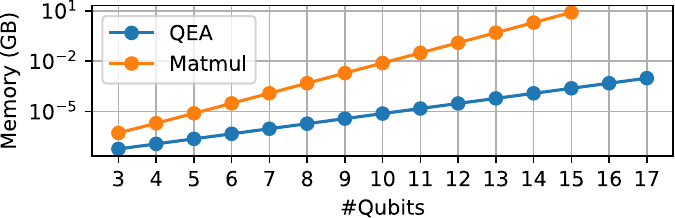}
            \caption{Comparison in memory usage (logarithm scale) between QEA and naive MM operation (Matmul).}
            \label{fig:mem_comparison}  
            \vspace{-5mm}
        \end{figure}

        \begin{figure*}[t]
            \centering
            \includegraphics[width=0.95\textwidth]{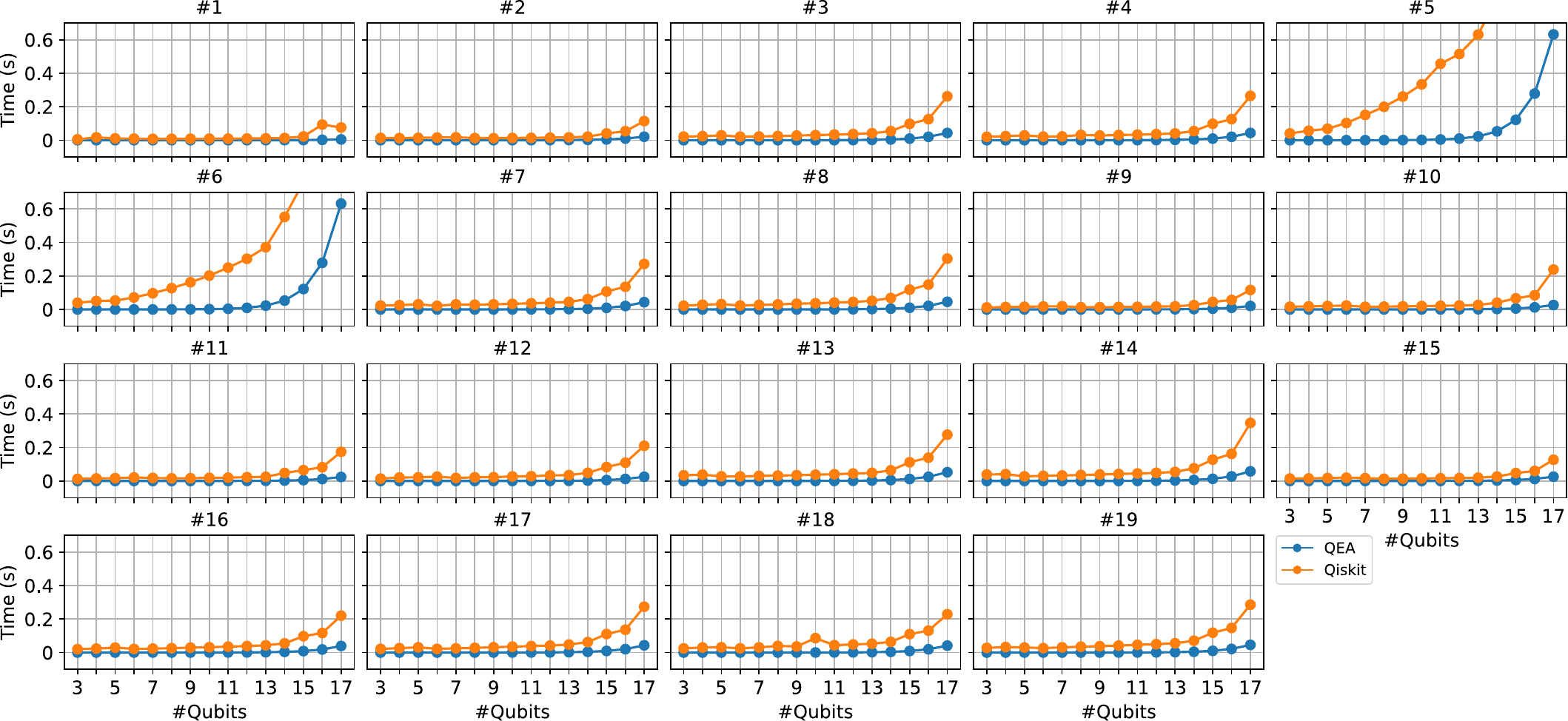}
            \caption{Comparison in execution time between QEA and a variety of random quantum circuits indexed from \#1 to \#19}
            \label{fig:execution_time_quanv_comparison}  
            \vspace{-5mm}
        \end{figure*}   
        The evaluated metrics include memory usage, fidelity, mean-square error (MSE), and execution time defined clearly in \cite{vu2024fqsun}. The fidelity and MSE between two state vectors range from 0 to 1 with fidelity of 1 or MSE of 0 meaning two vectors completely overlap and vice versa. The execution time on the CPU (Qiskit) is measured from constructing the circuit to receiving the final state while the execution time of QEA is measured by running QEA with a maximum frequency of 250 MHz to calculate the final state from the initial state. 
       
    \subsection{Comparison with powerful CPUs} \label{sec:ver_and_eval:benchmark_cpu:comparison}
        In this section, the Intel(R) Core(TM) i9-10940X CPU at 3.30GHz will be used to simulate quantum circuits supported by Qiskit. Its results are compared with the QEA's results. To demonstrate the effectiveness and optimization in memory management and allocation, an evaluation was conducted between QEA and the naive matrix multiplication (Matmul), as shown in Fig.~\ref{fig:mem_comparison}. The results show that as \#Qubits rises, QEA improves the amount of memory used. At 7 and 13 qubits, QEA surpasses the naive Matmul operation by factors ranging from approximately $\approx 10^2$ to $\approx 10^4$, demonstrating the efficiency of proposed memory management optimization.

        \begin{figure}[t]
            \centering
            \includegraphics[width=0.47\textwidth]{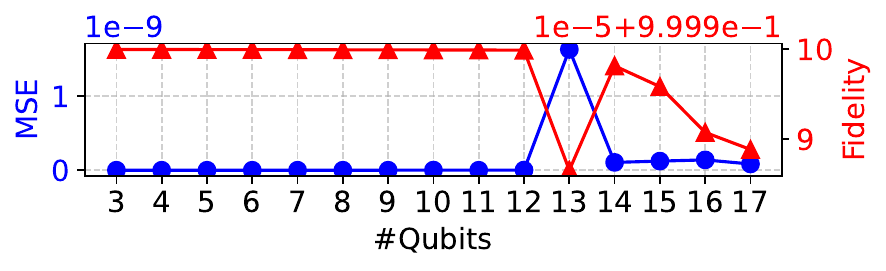}
            \caption{Average MSE and fidelity measurement between QEA and Qiskit for a set of parameterized quantum circuits indexed from 1 to 19.}
            \label{fig:mse_fidelity}  
            
        \end{figure}
        
\begin{figure}[t]
            \centering
            \includegraphics[width=0.47\textwidth]{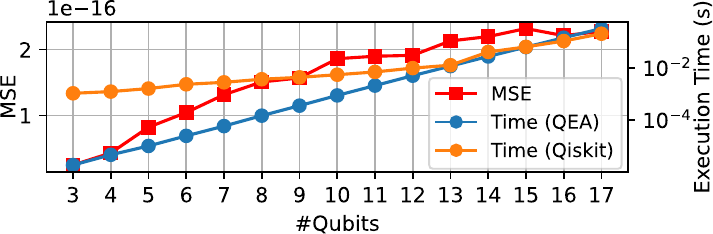}
            \caption{Comparison on QFT circuit. (a) Left y-axis: Mean square error between state vectors from Qiskit and QEA (b) Right y-axis: Execution time curve from Qiskit (Orange line) and QEA (Blue line).}
            \label{fig:qft}  
            
        \end{figure}

        In addition to memory usage, the comparison in terms of execution time between QEA and Qiskit with a set of parameterized quantum circuits indexed from 1 to 19 is shown in Fig.~\ref{fig:execution_time_quanv_comparison}. The QEA's results consistently achieve better execution times for circuits with a high \#Qubits, while its performance is comparable to Qiskit's for circuits with fewer \#Qubits. QEA also clearly shows its performance and effectiveness with applications using high \#Qubits while Qiskit cannot. For QFT circuits comparison in Fig.~\ref{fig:qft}, the execution time of Qiskit increases more slowly than that of QEA as \#Qubits increases, this phenomenon is explained in \cite{vu2024fqsun}. Qiskit does not use a state vector-based simulator in the case of a simple structure as QFT, then it can achieve better performance than QEA for high \#Qubits. Note that this acceleration technique from Qiskit can not work on general circuits.

        Fig.~\ref{fig:mse_fidelity} and Fig.~\ref{fig:qft} (left y-axis) show a detailed comparison of QEA and Qiskit on the accuracy aspect. The comparison reveals that there are no notable differences between QEA and Qiskit. As a result, QEA meets the MSE and fidelity standards, demonstrating its capacity as a dependable quantum emulator.
       
    \subsection{Comparison with FPGA-based works} \label{sec:ver_and_eval:FPGA_comparation}
        \begin{table*}[ht]
    		\centering
    		\renewcommand{\arraystretch}{1.2}
    		\resizebox{0.99\linewidth}{!}{\begin{threeparttable}
            \caption{Comparative analysis in post-implementation of QEA and existing FPGA-based emulators on QFT's performance.}
            \label{tab:qft}
            \begin{tabular}{|c|c|c|c|c|c|c|c|c|}
                \hline
                \textbf{Works} &
                \textbf{Device} & \textbf{Freq (MHz)}
                & \textbf{Reconfig$\star$} &
                \textbf{Precision} &
                \textbf{\#Qubits} &
                \begin{tabular}[c]{@{}c@{}}\textbf{Execution}  \textbf{time (s)}\end{tabular} &  
                \textbf{\#Gates $^{\dagger}$} &
                \textbf{NGS $^{\dagger\dagger}$} \\ 
                \hline
                \cite{silva2017fpga} & \begin{tabular}[c]{@{}c@{}}AMD Xilinx\\ Zynq-7000 \end{tabular} & 100 & \Checkmark & 32-bit FX & 6 & $1.15\times 10^{-4}$ & 10 & $1.8\times 10^{-7}$ \\ 
                \hline
                \cite{mahmud2020efficient} & \begin{tabular}[c]{@{}c@{}}Arria\\ 10AX115N4F45E3SG\end{tabular} & 233 & \Checkmark & 32-bit FP & 16 & $1.84\times10^{1}$ & 528 & $5.33\times 10^{-7}$ \\ 
                \hline
                \cite{hong2022quantum} & \begin{tabular}[c]{@{}c@{}}Xilinx XCKU115 \end{tabular} & 160 & \XSolidBrush & 16-bit FX & 16 & $2.70\times 10^{-1}$ & 136 & $3.03\times 10^{-8}$ \\ 
                \hline
                \cite{waidyasooriya2022scalable}& \begin{tabular}[c]{@{}c@{}} 2 $\times$ Intel Stratix 10\\ MX2100 \end{tabular} & 299 & \XSolidBrush & 32-bit FP & 30 & $4.47\times 10^0$ & 465 & $8.95\times 10^{-12}$ \\ 
                \hline
                \cite{liang2024pcq} & Xilinx XCVU9P & 233 & \XSolidBrush & 18-bit FX & 16 & $1.20\times 10^{-3}$ & - & - \\ 
                \hline
                \textbf{This work} & \textbf{AMD Alveo U280} & \textbf{250} & \Checkmark & \textbf{32-bit FX} & \textbf{17} & $\bm{3.29\times 10^{-1}}$ & \textbf{721} & $\bm{3.48\times 10^{-9}}$ \\ 
                \hline
            \end{tabular}
            \begin{tablenotes}
                \item[$^{\star}$] The quantum emulator is fixed with a application and \#Qubits. 
                \item[$^{\dagger}$] The number of gates ($\#\text{Gate}$) in this work is higher than other work due to the no-use of control-rotation gates.
                \item[$^{\dagger\dagger}$] The \textbf{N}ormalized \textbf{G}ate \textbf{S}peed (NGS) (s / (gate $\times$ amplitude)) = Execution time / (\#Gates $\times$ $2^{\#\text{Qubits}}$), smaller is better.
            \end{tablenotes}
            \label{tab:hardware_compare}
    		\end{threeparttable}}
            \vspace{-5mm}
    	\end{table*}
            
        This section compares QEA to other related studies on FPGA to provide a more comprehensive evaluation of performance and efficiency. The QFT technique was utilized to compare with other studies since it is widely used in the creation of hardware designs. Table~\ref{tab:hardware_compare} compares QEA's hardware results to other FPGA-based works \cite{liang2024pcq, mahmud2020efficient, hong2022quantum, waidyasooriya2022scalable, silva2017fpga} in terms of frequency, reconfigurable, \#Qubits, and normalized gate speed due to differences in supported qubits, precision, hardware resources, and the number of gates in quantum circuits.
        
        Table \ref{tab:hardware_compare} shows that QEA surpasses other works in terms of performance and reconfigurable. QEA outperforms \cite{liang2024pcq, mahmud2020efficient, hong2022quantum, silva2017fpga} in terms of maximum frequency, with increases ranging from 1.07$\times$ (250 vs. 233 MHz) to 2.5$\times$ (250 vs. 100 MHz). QEA also supports a broader range of algorithms, which increases its versatility and allows it to handle a higher \#Qubits. The NGS (Normalized Gate Speed) results for QEA additionally show a higher efficiency over previous works, with improvements ranging from 8.7$\times$ ($3.48 \times 10^{-9}$ vs. $3.03 \times 10^{-8}$ seconds) to 153.16$\times$ ($3.48 \times 10^{-9}$ vs. $5.33 \times 10^{-7}$ seconds). While QEA does not reach the best results in maximum frequency, supported qubits, and NGS when compared to \cite{waidyasooriya2022scalable}, it excels in providing better adaptability over a larger spectrum of quantum circuits, rather than being limited to QFT only. In summary, the results demonstrate QEA's superior features when compared to alternative FPGA-based works.
        
\section{Conclusion} \label{sec:concl}
    In summary, this paper introduced QEA, a quantum emulator, to address current challenges in achieving high flexibility, memory efficiency, low hardware resources, and high performance. QEA was presented with four key ideas: optimized memory allocation management, open PE, flexible ALU, and simplified CX swapper. The complete verification and evaluation confirmed the exceptional properties of QEA in terms of low MSE, high fidelity, less memory usage, short execution time, and high flexibility compared to other comparable works in supporting a variety of applications with varying \#Qubits from 3 to 17. In future work, QEA can be enhanced and integrated into a system using external data memories and multiple QEA cores to get higher performance with higher supported \#Qubits.

\section*{Acknowledgment}
This research is funded by the NAIST Scholar program. This work was supported by JST-ALCA-Next Program Grant Number JPMJAN23F5, Japan. The research has been partly executed in response to the support of JSPS, KAKENHI Grant No. 22H00515, Japan.

\bibliographystyle{IEEEtran}
\bibliography{references.bib}

\end{document}